\documentclass{iau} 
\usepackage{graphicx}

\usepackage{amsmath}
\usepackage{longtable}
\usepackage{supertabular}

\makeatletter
\@fleqnfalse
\@mathmargin\@centering
\makeatother

\usepackage{color}
\usepackage{natbib}

\newcommand{\RNum}[1]{\uppercase\expandafter{\romannumeral #1\relax}}
\newcommand{\arcsec}{$^{\prime\prime}$}

\newcommand{\uae}{\ensuremath{\bar{\mu}_{e}}}

\newcommand{\uaer}{\ensuremath{\bar{\mu}_{e,r}}}
\newcommand{\re}{\ensuremath{\bar{r}_{e}}}
\newcommand{\rer}{\ensuremath{\bar{r}_{e,r}}}

\newcommand{\MTObjects}{{\scshape MTObjects}}

\newcommand{\GALFIT}{{\scshape GALFIT}}
\newcommand{\SExtractor}{{\scshape SExtractor}}

\newcommand{\remin}{3.0}
\newcommand{\remax}{8.0}
\newcommand{\SBmin}{24.0}
\newcommand{\SBmax}{26.5}

\newcommand{\nmax}{2.5}

\newcommand{\SBsel}{\SBmin$\leq$\uaer$\leq$\SBmax}
\newcommand{\resel}{\remin\arcsec$\leq$\rer$\leq$\remax\arcsec}

\newcommand{\kidsarea}{180}
\newcommand{\UnmaskedArea}{39}

\newcommand{\UDGdensity}{8$\pm$3$\times10^{-3}$cMpc$^{-3}$}

\newcommand{\UDGdensityHI}{1.5$\pm$0.6$\times10^{-3}$cMpc$^{-3}$}
\newcommand{\UDGefficiency}{$\sim$0.8$\pm$0.2}

\title[UDGs in the field] {Observational Properties of Field UDGs: Colours and Number Densities}

\author[D. J. Prole, R. F. J. van der Burg, M. Hilker, J. I. Davies]   
{Daniel J. Prole$^{1,2,3,4}$, Remco F. J. van der Burg$^3$, Michael Hilker$^3$ \and Jonathan I. Davies$^4$}

\affiliation{
	$^1$Department of Physics \& Astronomy, Macquarie University, Sydney, NSW 2109, Australia
	\\ email: {\tt daniel.prole@mq.edu.au} \\[\affilskip]
	$^2$Australian Astronomical Optics, Macquarie University, Sydney, NSW 2109, Australia \\[\affilskip]
	$^3$European Southern Observatory, Karl-Schwarzschild-Str. 2, 85748 Garching bei M\"unchen, Germany  \\[\affilskip]
    $^4$School of Physics and Astronomy, Cardiff University, The Parade, Cardiff, CF243AA, UK }

\pubyear{2019}
\volume{355}  
\jname{The Realm of the Low Surface Brightness Universe}
\editors{D. Valls-Gabaud, I. Trujillo \& S. Okamoto, eds.}

\begin{document}
\maketitle

\begin{abstract}
	
While much of the focus around Ultra-Diffuse Galaxies (UDGs) has been given to those in galaxy groups and clusters, relatively little is known about them in less-dense environments. These isolated UDGs provide fundamental insights into UDG formation because environmentally driven evolution and survivability play less of a role in determining their physical and observable properties. We have recently conducted a statistical analysis of UDGs in the field using a new catalogue of sources detected in the deep Kilo-Degree Survey (KiDS) and Hyper Suprime-Cam Subaru Strategic Program (HSC-SSP) optical imaging surveys. Using an empirical model to assess our contamination from interloping sources, we show that a scenario in which cluster-like quiescent UDGs occupy a large fraction of the field UDG population is unlikely, with most being significantly bluer and some showing signs of localised star formation. We estimate an upper-limit on the total field abundance of UDGs of \UDGdensity\ within our selection range. The mass formation efficiency of UDGs implied by this upper-limit is similar to what is measured in groups and clusters, meaning that secular formation channels may significantly contribute to the overall UDG population.

\end{abstract}

\firstsection 
\section{Introduction}

\indent There are several theoretically-plausible formation channels that can explain the existence of ultra-diffuse galaxies (UDGs). These can be secular in nature, arising from high angular momentum of the parent halo and stellar feedback processes \citep[e.g.][]{Amorisco2016, DiCintio2017}, or instead driven by environmental effects such as ram pressure stripping and tidal heating \citep[e.g.][]{Collins2013, Yozin2015, Carleton2018, Jiang2018}. It is not currently known how important secular formation channels are in UDG production and this has important implications for galaxy formation models.

\indent One important measurement for understanding UDG formation observationally is whether UDGs are relatively more common in dense environments like galaxy clusters, which would imply that environmental processes play a positive role in UDG production, or whether there are relatively fewer, implying that they are more easily destroyed. \cite{vanderBurg2017} found that UDGs are relatively more common in higher-mass environments, but the community has not reached a consensus \citep{Pina2018}.

\indent UDGs in clusters are typically on the red sequence \citep{Koda2015, vanderBurg2016, Singh2019} but may be systematically bluer towards the outskirts of galaxy groups \citep{Roman2017b,  Alabi2018, Zaritsky2019} and in lower density environments \citep{Greco2018a, Greco2018b}, suggesting that interactions associated with a dense environment, possibly during cluster in-fall, can efficiently quench UDGs. 

\indent In this study, based on findings originally presented in \cite{Prole2019}, we add to the discussion by showing that UDGs in the field\footnote{We note that our working definition of the field is a representative piece of the Universe in which galaxy groups and clusters are included, but massive haloes naturally make up a relatively small fraction by mass.} are indeed systematically bluer than those in clusters, implying that quenching mechanisms are strongly linked with environmental density for these objects. We also provide the first-ever measurement of the total number density of the field UDG population and find that we cannot rule out a mass formation efficiency (i.e. the number of UDGs per unit total mass) similar to what it is in clusters, implying that secular formation channels may play a significant role in UDG production. 

\section{Data}
\label{section:data}

\noindent For source detection and structural parameter estimation, we use a \kidsarea\ deg$^{2}$ subset of data from the Astrowise \citep{McFarland2011} reduction of the Kilo-Degree Survey \citep[KiDS;][]{deJong2013, Kuijken2019} that overlaps with the GAMA spectroscopic survey \citep{Driver2011} equatorial fields. We use the $r$-band images for source detection because they are the deepest and have the best image quality. This is the same data used by \cite{vanderBurg2017} in their study of the UDG populations in galaxy groups and so we can make direct comparisons with their findings. Despite the GAMA overlap, redshift measurements are not available for most of our sources because they are generally much fainter than the limiting depth of GAMA at $r$=19.8 mag.

\indent While the KiDS $r$-band is sufficient to reach a limiting surface brightness of \uaer$\sim$26.5 (we quote all surface brightnesses in magnitudes per square arc-second), we additionally used the first data release of the overlapping Hyper-Suprime-Cam Subaru Strategic Program \citep[][]{Aihara2018} to measure colours. The HSC-SSP data is around 0.5 mag deeper than KiDS in the $r$-band, but has a reduced footprint (around a quarter) overlapping with the GAMA regions compared to the KiDS area we considered. This left us with $\sim$\UnmaskedArea\ deg$^{2}$ of unmasked data from which we measured the colours. 

\indent We did not use the HSC-SSP for detection because of its limited footprint and because of its relatively aggressive background subtraction. Furthermore, reliable stellar reflection halo \& artefact masks were available to us for KiDS imaging, which allowed for a reduction in the amount of false positives in our detection pipeline, critical for this work. For this analysis, we restricted ourselves to the $g$ and $r$ bands but this can be expanded in future studies. 

\section{Source Identification}
\label{section:Source Identification}

\noindent We used \MTObjects\ \citep{Teeninga2016} to construct a preliminary source catalogue from the KiDS $r$-band data, using default parameters. We applied a selection based on segment statistics produced by \MTObjects\ to select extended and LSB sources.

\indent We then used \GALFIT\ \citep{Peng2002} with KiDS point spread functions to fit single component S\'ersic profiles to each selected source, using \MTObjects\ segments as masks and segment statistics for initial parameter guesses. We constructed our UDG candidate catalogue using the \GALFIT\ S\'ersic parameters, specifically: \SBsel, \resel, $n\leq${\nmax}, where \uae, \re\ and $n$ are respectively the mean surface brightness within the effective radius, the circularised effective radius and S\'ersic index. Completeness estimates were obtained by means of artificial galaxy injections, described in the following section.

\section{Artificial Galaxy Injections}
\label{Artificial Galaxy Injection}

\begin{figure}
	\includegraphics[width=\linewidth]{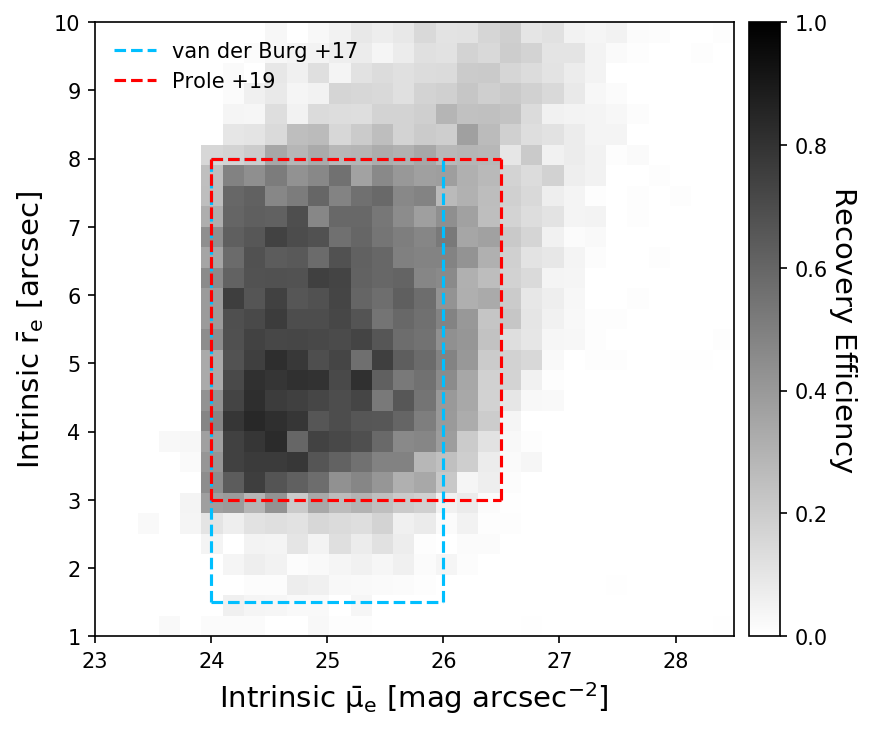}
	\centering
	\caption{The recovery efficiency of synthetic sources injected into the data as a function of circularised effective radius and mean surface brightness within the effective radius given our selection criteria. The red box indicates our selection criteria, while the blue box is that used in \protect\cite{vanderBurg2017}.}
	\label{figure:RE}
\end{figure}

\noindent Artificial galaxy injections were used to assess the statistical completeness of our source identification pipeline as a function of S\'ersic parameters \uae\ and \re; this is displayed in figure \ref{figure:RE}. 

\indent The artificial galaxies were made using \GALFIT\ and injected into the real data to produce augmented versions of each KiDS frame. The whole detection \& identification pipeline was run for each augmented frame, such that the recovered \GALFIT\ parameters could also be used to quantify our measurement errors. Overall, we injected $\sim$735,000 sources.

\indent We note that we have been able to probe $\sim$$0.5$ magnitudes deeper than \cite{vanderBurg2017} and this is likely due to our use of \MTObjects\ over \SExtractor, the latter of which is not designed to work in the LSB regime.

\section{Empirical Model}
\label{section:Emprical Model}

\noindent Since we do not know the distances to individual sources, we relied on an empirical model to statistically constrain the properties of the UDG candidates while accounting for interloping sources. Briefly, we modelled the UDGs with a fixed power-law size distribution and uniform surface brightness distribution \citep[cf.][]{vanderBurg2017}. We additionally used an empirical colour model based on dwarf galaxies in clusters \citep[cf.][]{Venhola2018}. We tested two colour models: a red/quiescent model, based on early-type dwarf galaxies, and a blue model, based on late-type dwarf galaxies.

\indent We argue that the dominant contribution to the interlopers in our UDG candidate catalogue are distant massive late-type galaxies that are cosmologically dimmed and thus satisfy our selection criteria. This is based on the fact that small/low-mass galaxies are statistically unlikely to be detected (given our recovery efficiency as a function of physical parameters and redshift) and that massive early-type galaxies general have high S\'ersic indices \citep{Vulcani2014}. We therefore chose to model the interloping population using canonical empirical relations such as the stellar mass function \citep{Baldry2012, Muzzin2013}, the stellar mass to size relation \citep{vanderWel2014} and an empirical colour model based on GAMA observations \citep{Taylor2011}, each pertaining specifically to late-type galaxies; we refer to this as the interloper model.

\indent We performed Monte Carlo sampling of this model (UDGs + interlopers) to generate mock observations, assuming that the galaxies trace the overall matter density of the Universe. We properly accounted for cosmological projection effects including $k$-corrections to derive apparent parameters. Our recovery efficiency measurement was used to assign a recovery probability to each mock galaxy, which was used to randomly select mock sources to construct a final mock observed galaxy catalogue. Such an approach naturally accounts for measurement uncertainties in our \GALFIT\ modelling.

\indent The number of Monte Carlo realisations was tuned so that the number of UDGs equals the expected value given a certain mass formation efficiency (the amount of mass is derived from the matter density of the Universe and the cosmological volume element). To begin with, we adopted a mass formation efficiency equal to that observed for UDGs in galaxy clusters \citep[c.f.][]{vanderBurg2017}.

\section{Results \& Discussion}
\label{section:Results}

\begin{figure*}
	\includegraphics[width=\linewidth]{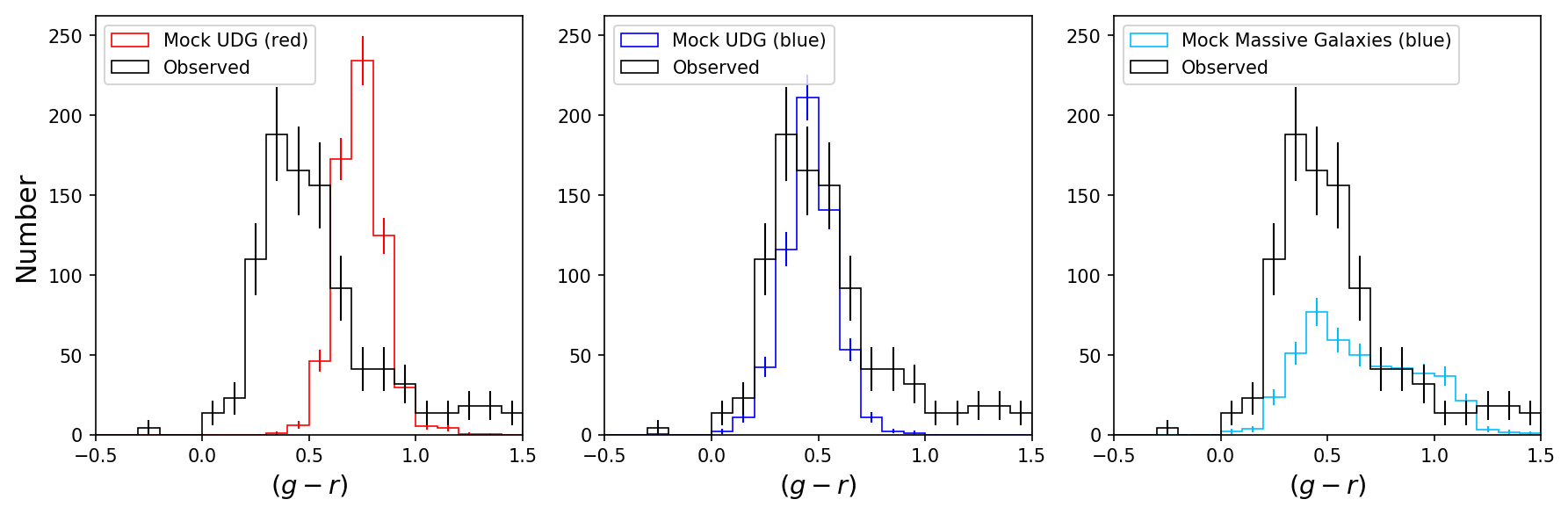}
	\centering
	\caption{Synthetic distributions of $(g-r)$ colour for our mock red UDG, blue UDG and interloper catalogues, weighted by the probability of observation, compared to the actual observed histogram. The absolute numbers are normalised to an area of 180 square degrees. The error-bars show the Poisson uncertainties in each bin. Colours are in the observed reference frame. We note that we include the effect of measurement error in our mock colours. It is clear that the red UDG model is not consistent with the observations, being much more consistent with the blue model.}
	\label{figure:grplot}
\end{figure*}

\begin{figure*}
	\includegraphics[width=\linewidth]{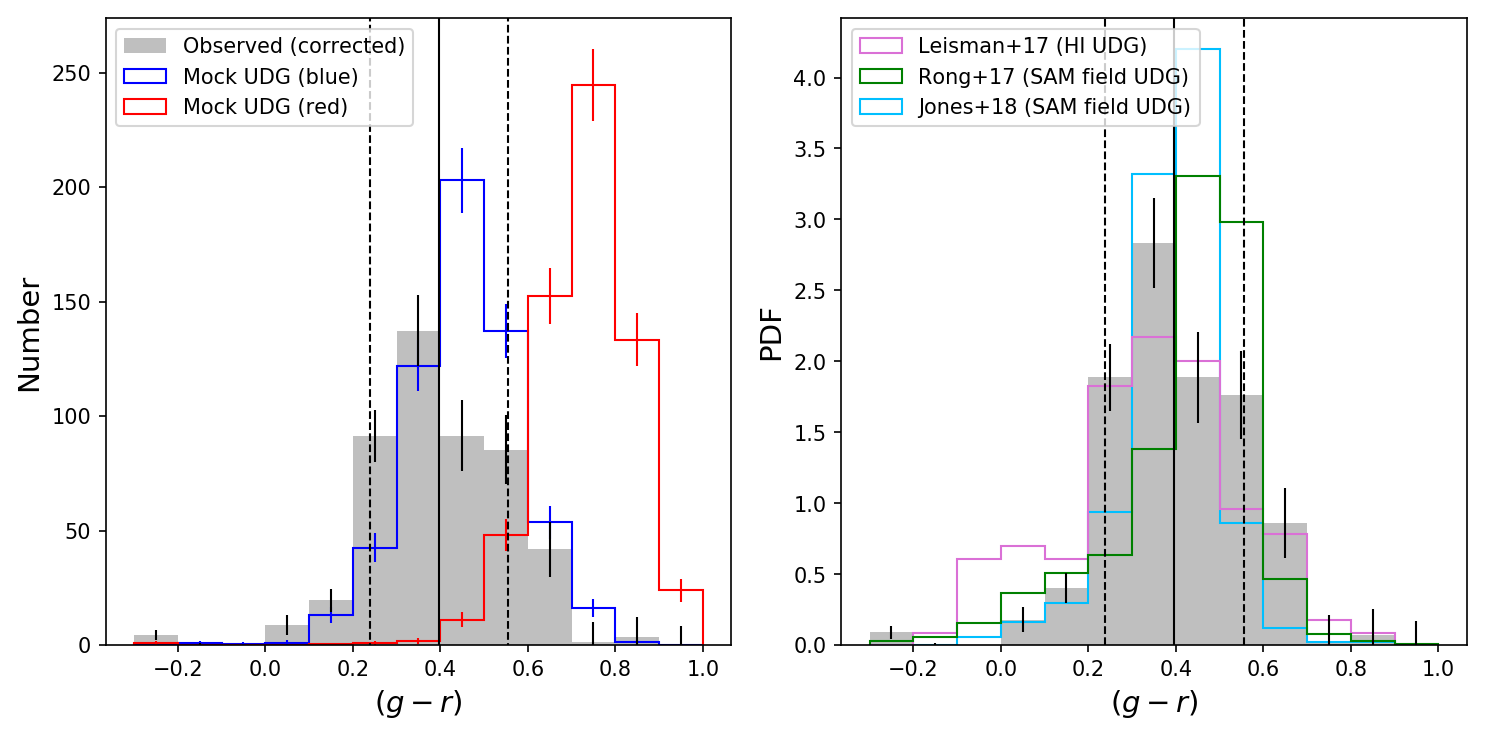}
	\centering
	\caption{The observed distribution of colour after subtracting the estimated contribution from massive blue galaxies (grey histogram). \textit{Left:} Comparison with the empirical red and blue UDG models from this study. We show the mean and 1$\sigma$ dispersion of the observations with the vertical lines. \textit{Right:} Normalised comparison with observations of H\RNum{1}-bearing field UDGs in the literature \protect\citep{Leisman2017}. and predictions from the semi-analytical models (SAM) of \protect\cite{Rong2017} and \protect\cite{Jones2018}. Poisson error-bars are shown. See text for discussion.}
	\label{figure:grplot_lit}
\end{figure*}

\noindent The result of sampling the red UDG model, the blue UDG model and the interloper model are shown consecutively in figure \ref{figure:grplot}. By comparing the colour distributions predicted from our mock catalogues with the real observations, it became clear that a scenario in which a large number (relative to that which would be expected given the assumed mass formation efficiency) of cluster-like quiescent UDGs existing in the field is unlikely (figure \ref{figure:grplot_lit}, first panel, where the corrected-observed histogram is obtained by statistically subtracting the interloper model from the observations).

\indent By comparing the absolute number count of our corrected-observed catalogue to that of our mock blue UDG catalogue, we find that we over-predict the number of UDGs by about 20\%. This implies that UDGs in the field form with an average mass formation efficiency \UDGefficiency\ times the cluster value. However, this remains an upper limit because we have only considered one population, namely massive late-type galaxies, in our interloper model. As discussed in \cite{Jones2018}, it is likely that our estimated number density of UDGs constitutes a negligible (of order one-percent) change to the galaxy stellar mass function in the dwarf galaxy regime.

\indent We also compared our observations with the sample of UDGs detected in H\RNum{1} by \cite{Leisman2017}. We found a reasonable qualitative agreement with the colour distribution of UDGs from their sample, indicating that the H\RNum{1}-rich UDGs have similar properties to the overall field UDG population. \cite{Jones2018} reported a number density of H\RNum{1}-rich UDGs of \UDGdensityHI, approximately one-fifth of our upper-limit measurement of the overall UDG field population.

\indent Finally, we compared our results to the semi-analytic models (SAMs) of \cite{Rong2017} and \cite{Jones2018}, finding a qualitative agreement in colour. However, we found that the SAM used by \cite{Jones2018} over-predicts the number of field UDGs compared to our observations by a factor of $\sim$2.

\indent To summarise, this study suggests that UDGs in the field are systematically bluer than those in denser environments and likely harbour younger stellar populations, with some showing clear signs of active star formation. This is observational evidence that environmental effects play an important role in quenching star formation in such galaxies. The upper-limit mass formation efficiency we have estimated is not sufficient to rule out the idea that secular formation channels are responsible for producing a high fraction of UDGs.

\begin{discussion}
	
	\discuss{A. Di Cintio}{Do your models follow the main UDG relations\,?}
	\discuss{D. Prole}{Yes, our empirical model relies on scaling relations found for UDGs and dwarf galaxies in clusters. Testing these assumptions will be a key objective of future work, for example, constraining the intrinsic size distribution of UDGs in the field.}
	
	\discuss{R. Taylor}{What is the field number density of these UDGs as compared with the ones in clusters, and
		UDGs vs the normal population\,?}
	\discuss{D. Prole}{We find a roughly equivalent mass formation efficiency with those in clusters, although this is an upper limit. Since most of the mass in the Universe is not in clusters, it may be that most UDGs in the Universe are isolated. Looking at the stellar mass function for dwarf galaxies and comparing with our estimated number density, field UDGs probably comprise of order $\sim$1\% of the dwarf galaxy population.}
	
	\discuss{P.A. Duc}{Could the problem of  not having proper distances for UDGs bias the estimate of their
		number densities\,?}
	\discuss{D. Prole}{Our empirical model only accounts for massive late-type galaxy interlopers, so it is possible that other types of galaxies like small and nearby dwarf galaxies and perhaps intermediate-mass early-types also make it into our sample. That is why our quoted number density is an upper-limit estimate only. However, our modelling suggests that other types of interloping galaxy are in the minority compared to massive late-types.}
	
\end{discussion}


\begin{thebibliography}
		   
\bibitem[Aihara \etal (2018)]{Aihara2018}{Aihara H. \etal} 2018, \textit{Publications of the Astronomical Society of Japan}, 70, S8
\bibitem[Alabi \etal (2018)]{Alabi2018}{Alabi A. \etal} 2018, \textit{MNRAS}, 479, 3308-3318
\bibitem[Amorisco \& Loeb (2016)]{Amorisco2016}{Amorisco N.~C., Loeb A.} 2016, \textit{MNRAS}, 459, L51-L55
\bibitem[Baldry \etal (2012)]{Baldry2012}{Baldry I.~K. \etal} 2012, \textit{MNRAS}, 421, 621-634
\bibitem[Carleton \etal (2018)]{Carleton2018}{Carleton T., Errani R., Cooper M., Kaplinghat M., Pe\~narrubia J.} 2018, arXiv eprints, arXiv:1805.06896
\bibitem[Collins \etal (2013)]{Collins2013}{Collins M.~L.~M. \etal} 2013, \textit{The Astrophysical Journal}, 768, 172
\bibitem[Di Cintio \etal (2017)]{DiCintio2017}{Di Cintio A., Brook C.~B., Dutton A.~A., Macci\`o A.~V., Obreja A., Dekel A.} 2017, \textit{MNRAS}, 466, L1-L6
\bibitem[Driver \etal (2011)]{Driver2011}{Driver S.~P. \etal} 2011, \textit{MNRAS}, 413, 971-995
\bibitem[Greco \etal (2018)]{Greco2018a}{Greco J.~P. \etal} 2018, \textit{The Astrophysical Journal}, 857, 104
\bibitem[Greco \etal (2018b)]{Greco2018b}{Greco J. P., Goulding A. D., Greene J. E., Strauss M. A., Huang S., Kim Ji H., Komiyama Y.} 2018, \textit{The Astrophysical Journal}, 866, 112
\bibitem[Jiang \etal (2018)]{Jiang2018}{Jiang F., Dekel A., Freundlich J., Romanowsky A. J., Dutton A., Maccio A., Di Cintio A.} 2018, arXiv eprints, arXiv:1811.10607
\bibitem[Jones \etal (2018)]{Jones2018}{Jones M.~G. \etal} 2018, \textit{Astronomy and Astrophysics}, 614, A21
\bibitem[Koda \etal (2015)]{Koda2015}{Koda J., Yagi M., Yamanoi H., Komiyama Y.} 2015, \textit{The Astrophysical Journal Letters}, 807, L2
\bibitem[Kuijken \etal (2019)]{Kuijken2019}{Kuijken K. \etal} 2019, \textit{Astronomy and Astrophysics}, 625, A2
\bibitem[Leisman \etal (2017)]{Leisman2017}{Leisman L. \etal} 2017, \textit{The Astrophysical Journal}, 842, 133
\bibitem[McFarland \etal (2011)]{McFarland2011}{McFarland J.~P., Verdoes-Kleijn G., Sikkema G., Helmich E.~M., Boxhoorn D.~R., Valentijn E.~A.} 2011, arXiv eprints, arXiv:1110.2509
\bibitem[Muzzin \etal (2013)]{Muzzin2013}{Muzzin A. \etal} 2013, \textit{The Astrophysical Journal}, 777, 18
\bibitem[Peng \etal (2002)]{Peng2002}{Peng C.~Y., Ho L.~C., Impey C.~D., Rix H.-W.} 2002, \textit{The Astronomical Journal}, 124, 266-293
\bibitem[Mancera \etal (2018)]{Pina2018}{Mancera P. P.~E., Peletier R.~F., Aguerri J.~A.~L., Venhola A., Trager S., Choque C. N.} 2018, \textit{MNRAS}, 481, 4381-4388
\bibitem[Prole \etal (2019)]{Prole2019}{Prole D.~J., van der Burg R.~F.~J., Hilker M., Davies J.~I.} 2019, \textit{MNRAS}, 488, 2143-2157
\bibitem[Rom\'an \& Trujillo (2017)]{Roman2017b}{Rom\'an J., Trujillo I.} 2017, \textit{MNRAS}, 468, 4039-4047
\bibitem[Rong \etal (2017)]{Rong2017}{Rong Y. \etal} 2017, \textit{MNRAS}, 470, 4231-4240
\bibitem[Singh \etal (2019)]{Singh2019}{Singh P. R., Zaritsky D., Donnerstein R., Spekkens K.} 2019, \textit{The Astronomical Journal}, 157, 212
\bibitem[Taylor \etal (2011)]{Taylor2011}{Taylor E.~N. \etal} 2011, \textit{MNRAS}, 418, 1587-1620
\bibitem[Teeninga \etal (2016)]{Teeninga2016}{Teeninga P., Moschini U., Trager S. C., Wilkinson M. H.} 2016\textit{International Symposium on Mathematical Morphology and Its Applications to Signal and Image Processing} 157–168
\bibitem[Venhola \etal (2018)]{Venhola2018}{Venhola A. \etal} 2018, \textit{Astronomy and Astrophysics}, 620, A165
\bibitem[Vulcani \etal (2014)]{Vulcani2014}{Vulcani B. \etal} 2014, \textit{MNRAS}, 441, 1340-1362
\bibitem[Yozin \& Bekki (2015)]{Yozin2015}{Yozin C., Bekki K.} 2015, \textit{MNRAS}, 452, 937-943
\bibitem[Zaritsky \etal (2019)]{Zaritsky2019}{Zaritsky D. \etal} 2019, \textit{The Astrophysical Journal Supplement Series}, 240, 1
\bibitem[de Jong \etal (2013)]{deJong2013}{de Jong J.~T.~A., Verdoes K. G.~A., Kuijken K.~H., Valentijn E.~A.} 2013, \textit{Experimental Astronomy}, 35, 25-44
\bibitem[van der Burg \etal (2016)]{vanderBurg2016}{van der Burg R.~F.~J., Muzzin A., Hoekstra H.} 2016, \textit{Astronomy and Astrophysics}, 590, A20
\bibitem[van der Burg \etal (2017)]{vanderBurg2017}{van der Burg R.~F.~J. \etal} 2017, \textit{Astronomy and Astrophysics}, 607, A79
\bibitem[van der Wel \etal (2014)]{vanderWel2014}{van der Wel A. \etal} 2014, \textit{The Astrophysical Journal}, 788, 28


\end{thebibliography}
\end{document}